\documentclass[conference]{IEEEtran}
\IEEEoverridecommandlockouts
\usepackage{cite}
\usepackage{amsmath,amssymb,amsfonts}
\usepackage{graphicx}
\usepackage{bbm}
\usepackage{booktabs}
\usepackage[caption=false,font=footnotesize]{subfig}
\usepackage{balance}
\usepackage{xcolor}
\graphicspath{{Figures/}}
\def\BibTeX{{\rm B\kern-.05em{\sc i\kern-.025em b}\kern-.08em
    T\kern-.1667em\lower.7ex\hbox{E}\kern-.125emX}}

\begin{document}

\title{3D Protraction Technique for Anchor-Free, Range-Free Wireless Localization in Industrial Warehouses via Multi-Level Binary Reachability\\
\thanks{This work was supported in part by the McGill University Vadasz Scholars Program.}}

\author{\IEEEauthorblockN{Peyman Moeini}
\IEEEauthorblockA{\textit{Department of Electrical and Computer Engineering} \\
\textit{McGill University}\\
Montreal, Canada \\
peyman.moeini@mail.mcgill.ca}
\and
\IEEEauthorblockN{Mark Coates}
\IEEEauthorblockA{\textit{Department of Electrical and Computer Engineering} \\
\textit{McGill University}\\
Montreal, Canada \\
mark.coates@mcgill.ca}}

\maketitle

\begin{abstract}
Three-dimensional (3D) localization is difficult in industrial facilities because fixed-node placement is constrained by hardware, power, networking, installation, and maintenance costs. This paper presents the 3D Protraction Technique (3D-PT), an anchor-free, range-free method that converts reciprocal packet detections at ordered transmit-power levels into reliability-weighted ordinal distance relations. It estimates a relative 3D configuration without Received Signal Strength Indicator (RSSI) magnitudes, estimated ranges, surveyed anchors, or prior coordinates. In a simulated multi-height warehouse, four transmit-power levels reduce mean network-diameter-normalized RMSE by 16.0\% relative to MDS-MAP(P) in both small- and medium-network regimes, and by 23.9\% and 34.0\% relative to CATL-reproduction, respectively. Accuracy largely saturates beyond four power levels. The results show that multi-level binary reachability provides useful geometric information beyond a single maximum-power connectivity graph.
\end{abstract}


\section{Introduction}
\label{sec:introduction}

Wireless localization supports industrial automation, asset tracking, inventory management, and production monitoring \cite{Laoudias2018,Park2018}. In warehouses, however, metal racks, machinery, stored material, and walls create difficult propagation, while mounting locations, electrical power, and network access constrain fixed-node placement \cite{Gong2016,NoorARahim2022,Gong2018,Macoir2019}. The number and placement of fixed nodes therefore affect feasibility, installation, configuration, and maintenance cost \cite{moeini2025RTLS,Bendavid2024,Lazaro2021}. Systems based on Received Signal Strength Indicator (RSSI), time, or angle measurements can provide metric information but may require calibration, synchronization, specialized hardware, or surveyed reference coordinates \cite{Rathnayake2023,Arigye2022}. These requirements are especially restrictive in three-dimensional (3D) industrial environments with limited placement flexibility and poor reference geometry \cite{Kanhere2018,Liu2019,KhalafAllah2021}.

Anchor-free, range-free (AF--RF) localization avoids both surveyed coordinates and metric ranging \cite{Amundson2009,Kaur2025,Rajan2019}. Existing AF--RF methods generally infer geometry from a single connectivity graph, so their information content depends strongly on node density and graph structure \cite{Roman2026}. Using multiple transmit-power levels provides more information: pairs detected at lower power levels are likely closer than pairs detected only at higher levels. This evidence is ordinal rather than metric and can be obtained from binary packet outcomes without exposing received-power values to the localization algorithm.

This paper extends the two-dimensional (2D) Protraction Technique (PT) \cite{Moeini2026} to 3D industrial localization. The proposed 3D-PT (i) converts each reciprocal detection signature into an ordinal proximity score and an explicit consistency-based reliability, (ii) constructs clearly defined local and global candidate relations and selects reliable relations before coordinate optimization, and (iii) estimates a relative 3D node configuration using robust, scale-normalized ordinal penalties. We compare 3D-PT with the patch based MDS-MAP(P) algorithm and Connectivity-based and Anchor-free Three-dimensional Localization (CATL) reproduction in a multi-height warehouse and quantify the computation accuracy trade-off as the number of reachability levels increases.

\section{Related Work}
\label{sec:related_work}

Connectivity-based multidimensional-scaling methods form a prominent
family of AF--RF localization techniques \cite{Saeed2019}. MDS-MAP estimates relative node positions from connectivity-derived distances, while MDS-MAP(P) improves performance in irregular networks by constructing and merging overlapping local maps \cite{Priyantha2005,Zhao2018}. Its performance nevertheless depends on sufficiently accurate local embeddings: low connectivity degrades the local maps, and errors can propagate through patch merging \cite{Shang2004}. The appropriate patch size also depends on network density and topology, which is problematic when the available infrastructure-node population is intentionally small.

CATL specifically addresses anchor-free connectivity-based localization in large 2D and 3D networks with concave regions \cite{Tan2013}. It reduces strongly bent shortest-path effects through notch-aware multilateration and selects four internal references for 3D reconstruction. Its hop-based estimates, however, assume a relatively dense and approximately uniform node distribution. Nonuniform or sparse regions can provide too few reliable paths, and some connectivity graphs do not admit a unique 3D realization \cite{Tan2013,Han2020}.

The earlier 2D-PT introduced multi-level binary reachability as ordinal geometric evidence \cite{Moeini2026}. The present work extends this principle to $\mathbb R^3$ and adds global pair-order relations to the reference-centered local comparisons. Unlike MDS-MAP(P) and CATL, 3D-PT uses observations across the ordered transmit power levels rather than reducing them immediately to one maximum-power connectivity graph.

\section{System Model and Objective}
\label{sec:system_model}

Consider $n$ fixed nodes at unknown positions
\begin{equation}
\mathbf{x}_i=[x_i,y_i,z_i]^{\mathsf T}\in\mathbb R^3,
\qquad i\in\mathcal V=\{1,\ldots,n\}.
\label{eq:trueposition}
\end{equation}
No surveyed anchors, ranges, RSSI magnitudes, timing, angular measurements, floor labels, or obstacle geometry are supplied to the localization algorithm.

All nodes share the ordered transmit-power set $\mathcal P_{\rm tx}=\{p_1<\cdots<p_K\}$. Let $h\in\{1,\ldots,K\}$ index the transmit-power level, where level $h$ uses power $p_h$. At this level, node $i$ transmits $M_h$ packets.
For packet index $m\in\{1,\ldots,M_h\}$, define $b_{ij}^{(m)}(h)\in\{0,1\}$ as the binary detection outcome at node $j$: $b_{ij}^{(m)}(h)=1$ if node $j$ detects the $m$th packet transmitted by node $i$ at power $p_h$, and $b_{ij}^{(m)}(h)=0$ otherwise. The empirical detection rate and directed binary state are
\begin{equation}
\hat q_{ij}(h)=\frac{1}{M_h}\sum_{m=1}^{M_h}b_{ij}^{(m)}(h),\qquad
a_{ij}^{(h)}=\mathbbm 1[\hat q_{ij}(h)\geq\theta_{\rm det}].
\label{eq:directedstate}
\end{equation}

where $\theta_{\rm det}\in[0,1]$ is the packet-detection-rate threshold. Also, $a_{ij}^{(h)}$ denotes reachability from transmitter $i$ to receiver $j$ at level $h$, while $a_{ji}^{(h)}$ denotes the corresponding reverse-direction reachability from $j$ to $i$. Reciprocal reachability is $r_{ij}^{(h)}=a_{ij}^{(h)}a_{ji}^{(h)}$, and the pair signature is
\begin{equation}
\mathbf r_{ij}=[r_{ij}^{(1)},\ldots,r_{ij}^{(K)}]^{\mathsf T}\in\{0,1\}^K.
\label{eq:signature}
\end{equation}
Increasing power is assumed to increase reachability statistically, but fading, shadowing, and non-reciprocity can cause individual reversals. We require only that the reciprocal graph at $p_K$ be connected; disconnected components cannot be positioned relative to one another from connectivity alone.

The objective is to estimate $\hat{\mathbf X}$ such that its pairwise distances agree as closely as possible with the ordinal information in $\{\mathbf r_{ij}\}$. With neither anchors nor metric ranges, the result is defined only up to translation, rotation/reflection, and uniform scale \cite{Zhang2014}.

\section{3D Protraction Technique}
\label{sec:method}

A \emph{trial embedding} is a temporary 3D configuration of the nodes considered during optimization, where each node $i$ is assigned a candidate coordinate $\mathbf y_i\in\mathbb R^3$. These coordinates are collected row-wise in
\begin{equation}
\mathbf Y=[\mathbf y_1,\ldots,\mathbf y_n]^{\mathsf T}
\in\mathbb R^{n\times3},
\qquad
d_{ij}(\mathbf Y)=\|\mathbf y_i-\mathbf y_j\|_2,
\label{eq:embeddingdistance}
\end{equation}
where $d_{ij}(\mathbf Y)$ is the Euclidean distance between nodes $i$ and $j$ in the trial configuration. Thus, node pairs are not assigned separate coordinates; their relationship is represented by the distances induced by the common configuration $\mathbf Y$.

3D-PT constructs ordinal relations among these distances and evaluates each trial configuration using the scalar objective $\mathcal L(\mathbf Y)$, defined below. The objective penalizes violations of the selected local and global distance orderings together with a scale-normalization term. Thus, 3D-PT seeks a configuration
\begin{equation}
\mathbf Y^\star
\in
\underset{\mathbf Y\in\mathbb R^{n\times3}}{\operatorname{arg\,min}}
\mathcal L(\mathbf Y),
\label{eq:optimalembedding}
\end{equation}
with smaller values of $\mathcal L(\mathbf Y)$ indicating greater agreement with the selected ordinal relations. The optimized relative configuration is reported as $\hat{\mathbf X}$.

\subsection{Binary Protraction Scores}

Ideally, a link detected at a given transmit-power level should remain detectable at higher levels. Wireless variability can violate this ordering; for example, a binary signature may contain $1,0$ at two consecutive increasing power levels. We refer to such a $1$-to-$0$ transition as a \emph{power-order reversal}. Each reciprocal signature is therefore projected onto the set of nondecreasing sequences:
\begin{equation}
\mathbf z_{ij}=
\underset{\substack{\boldsymbol\phi\in[0,1]^K\\
\phi_1\leq\cdots\leq\phi_K}}
{\operatorname{arg\,min}}
\|\boldsymbol\phi-\mathbf r_{ij}\|_2^2,
\label{eq:isotonic}
\end{equation}
where $\mathbf z_{ij}=[z_{ij}^{(1)},\ldots,z_{ij}^{(K)}]^{\mathsf T}$ is the
isotonic-regression estimate of $\mathbf r_{ij}$.

The projected sequence is converted back to binary form, and its average
defines the ordinal proximity score:
\begin{equation}
\tilde r_{ij}^{(h)}=
\mathbbm 1[z_{ij}^{(h)}\geq\tfrac12],
\qquad
s_{ij}=\frac1K\sum_{h=1}^K\tilde r_{ij}^{(h)}.
\label{eq:score}
\end{equation}
Thus, $s_{ij}\in[0,1]$ is the fraction of projected power levels at which nodes $i$ and $j$ are reciprocally reachable. A pair becoming reachable at a lower power level receives a larger score. Equal weighting avoids assuming that the power-level index $h$ is linearly related to physical power $p_h$; the score is ordinal and is not interpreted as a range estimate. Setting $s_{ii}=0$ gives the score vector $\mathbf s_i=[s_{i1},\ldots,s_{in}]^{\mathsf T}$ for reference node $i$.

\subsection{Pair Reliability}

The score $s_{ij}$ does not preserve two observation-quality effects: directional disagreement and power-order reversals. We therefore define
\begin{align}
c_{ij}^{\rm rec}
&=
1-\frac1K\sum_{h=1}^{K}
|a_{ij}^{(h)}-a_{ji}^{(h)}|,
\label{eq:recconsistency}\\
\nu_{ij}
&=
\begin{cases}
\dfrac{1}{K-1}\displaystyle\sum_{h=1}^{K-1}
\mathbbm 1[a_{ij}^{(h+1)}<a_{ij}^{(h)}],
& K\geq2,\\[2mm]
0, & K=1,
\end{cases}
\label{eq:reversalrate}\\
c_{ij}^{\rm mon}
&=
1-\tfrac12(\nu_{ij}+\nu_{ji}),
\qquad
\kappa_{ij}
=
(c_{ij}^{\rm rec}c_{ij}^{\rm mon})^{\beta_r}.
\label{eq:reliability}
\end{align}

The exponent $\beta_r>0$ controls how strongly reciprocity and monotonicity  inconsistencies reduce the pair reliability; its value is specified in Table~\ref{tab:parameters}.

Also, $c_{ij}^{\rm rec}\in[0,1]$ measures directional agreement, $\nu_{ij}$ is the fraction of adjacent power levels exhibiting a $1$-to-$0$ reversal in direction $i\!\rightarrow\!j$, $c_{ij}^{\rm mon}\in[0,1]$ measures monotonic consistency, and $\kappa_{ij}\in[0,1]$ is the resulting pair reliability. 

\subsection{Local Candidate Relations and Penalties}

For reference node $i$, let
$\pi_i(1),\ldots,\pi_i(n-1)$ denote a permutation of the other node indices ordered by decreasing score:
\[
s_{i\pi_i(1)}\geq\cdots\geq s_{i\pi_i(n-1)}.
\]
Thus, $\pi_i(r)$ is the node having rank $r$ relative to reference node $i$.

A \emph{candidate local relation} $(i,j,k)$ proposes that $j$ is closer to reference node $i$ than $k$. Candidates are generated using three deterministic rank comparisons: \emph{close--close} compares two high score nodes, \emph{close--far} compares a high-score node with a low score node, and \emph{stratified} compares nodes drawn from approximately uniformly spaced ranks across the full ordering. These rules determine only which candidate triplets are considered; their exact rank counts are specified in Section~\ref{sec:experimental}.

The deterministic rank counts are given in Section~\ref{sec:experimental}.
Let $\mathcal C_i^{\rm loc}$ denote the local candidate set generated for reference node $i$, and define
\[
\mathcal C^{\rm loc}
=
\bigcup_{i\in\mathcal V}\mathcal C_i^{\rm loc}.
\]

For each candidate $(i,j,k)\in\mathcal C^{\rm loc}$, define the score gap
\begin{equation}
\Delta s_{ijk}=s_{ij}-s_{ik},
\label{eq:localscoregap}
\end{equation}
which measures the strength of the proposed ordinal preference. 

We retain only candidates having a positive score gap and sufficiently reliable constituent links. The selected local-relation set is

\begin{equation}
\mathcal T=
\left\{(i,j,k)\in\mathcal C^{\rm loc}:
\Delta s_{ijk}>0,\;
\kappa_{ij}\geq\kappa_{\min},\;
\kappa_{ik}\geq\kappa_{\min}
\right\},
\label{eq:selectedlocal}
\end{equation}
Thus, $\mathcal T$ contains the local ordinal relations used in the optimization, and $\kappa_{\min}$ is the minimum accepted pair reliability. Selection is performed from the observations before optimizing $\mathbf Y$; neither the true coordinates nor the current trial embedding is used.

To limit the influence of strongly violated relations, we use the bounded-influence robust softplus
\begin{equation}
\rho(t)
=
c_\rho
\log\!\left[
1+\frac{\log(1+e^t)}{c_\rho}
\right],
\label{eq:robustsoftplus}
\end{equation}
where $t$ is the scalar penalty argument and $c_\rho>0$ is the
robustification parameter.

Each $(i,j,k)\in\mathcal T$ supplies noisy evidence for
$d_{ij}(\mathbf Y)<d_{ik}(\mathbf Y)$ and contributes the soft penalty
\begin{equation}
\ell_{ijk}(\mathbf Y)=
\rho\!\left(
\frac{
d_{ij}(\mathbf Y)-d_{ik}(\mathbf Y)+\mu_{ijk}
}{\tau}
\right),
\label{eq:localpenalty}
\end{equation}
where $\tau>0$ controls penalty smoothness and $\mu_{ijk}>0$ is the
requested ordinal separation:
\begin{align}
\mu_{ijk}
&=
\mu_{\rm loc}+\chi_t\Delta s_{ijk},
\label{eq:localmargin}\\
\tilde w_{ijk}
&=
\xi_t(\Delta s_{ijk})^{\gamma_{\rm loc}}
\big(\sqrt{\kappa_{ij}\kappa_{ik}}\big)^{\eta_{\rm loc}}.
\label{eq:localweight}
\end{align}

Here, $\tilde w_{ijk}$ is an unnormalized relation weight: larger score gaps and more reliable constituent links give the relation greater influence.
The relation class $t$ determines $\chi_t=\chi_{\rm cc}$ for close--close relations and $\chi_t=\chi_{\rm cf}$ otherwise; similarly, $\xi_t=\xi_{\rm strat}$ for stratified relations and $\xi_t=1$ otherwise. 
The exponents $\gamma_{\rm loc}>1$ and $\eta_{\rm loc}>1$ control the emphasis on larger score gaps and more reliable links, respectively. The geometric mean $\sqrt{\kappa_{ij}\kappa_{ik}}$ combines the reliabilities of the two links.

The unnormalized weight is normalized to unit mean over $\mathcal T$:
\begin{equation}
w_{ijk}
=
\frac{\tilde w_{ijk}}
{\frac{1}{|\mathcal T|}
\sum_{(a,b,c)\in\mathcal T}\tilde w_{abc}}.
\label{eq:localweightnorm}
\end{equation}
The normalized weight $w_{ijk}$ multiplies
$\ell_{ijk}(\mathbf Y)$ in the local term of the overall objective in
\eqref{eq:objective}.

\subsection{Global Relations and Ordinal Embedding}

Local triples compare distances sharing the same reference node. To add configuration-wide information, 3D-PT also forms a candidate set $\mathcal Q$ of comparisons between unordered node pairs ranked by decreasing reliable score. For $(i,j,u,v)\in\mathcal Q$, define
\begin{equation}
\Delta s_{ijuv}^{(g)}=s_{ij}-s_{uv}.
\end{equation}
The selected global relation set is
\begin{equation}
\mathcal G=
\left\{(i,j,u,v)\in\mathcal Q:
\Delta s_{ijuv}^{(g)}\geq\delta_g,\;
\kappa_{ij}\geq\kappa_g,\;
\kappa_{uv}\geq\kappa_g
\right\},
\label{eq:selectedglobal}
\end{equation}
where $\delta_g>0$ is the minimum accepted global score gap and $\kappa_g$ is the minimum reliability required for each pair participating
in a global relation.

A selected quadruple proposes
$d_{ij}(\mathbf Y)<d_{uv}(\mathbf Y)$ and uses
\begin{equation}
\bar\mu_{ijuv}
=
\mu_g+\chi_g\Delta s_{ijuv}^{(g)},
\qquad
\tilde{\bar w}_{ijuv}
=
(\Delta s_{ijuv}^{(g)})^{\gamma_g}
\sqrt{\kappa_{ij}\kappa_{uv}},
\label{eq:globalmarginweight}
\end{equation}
where $\bar\mu_{ijuv}$ is its ordinal margin, $\tilde{\bar w}_{ijuv}$ is its unnormalized weight, and $\gamma_g>0$ controls the emphasis on larger global score gaps. The normalized global weights $\bar w_{ijuv}$ have unit mean over $\mathcal G$. Neither local nor global score gaps are interpreted as metric distances.

The overall aggregate penalty, or objective function,
$\mathcal L(\mathbf Y)$, is
\begin{align}
\mathcal L(\mathbf Y)={}&
\sum_{(i,j,k)\in\mathcal T}
w_{ijk}\ell_{ijk}(\mathbf Y)
\nonumber\\
&+
\lambda_G
\sum_{(i,j,u,v)\in\mathcal G}
\bar w_{ijuv}
\rho\!\left(
\frac{
d_{ij}(\mathbf Y)-d_{uv}(\mathbf Y)+\bar\mu_{ijuv}
}{\tau}
\right)
\nonumber\\
&+
\lambda_S[\bar d^2(\mathbf Y)-1]^2,
\label{eq:objective}
\end{align}
where $\lambda_G$ weights the global ordinal term relative to the local term, and $\lambda_S$ weights the scale-normalization penalty. The scale statistic is
\begin{equation}
\bar d^2(\mathbf Y)
=
\frac{2}{n(n-1)}
\sum_{i<j}d_{ij}^2(\mathbf Y),
\label{eq:scale}
\end{equation}
and the bounded-influence robust softplus is

The scale term removes uniform-scale ambiguity. Optimization uses multiple spectral initializations followed by coarse-to-fine Adam refinement with
gradient clipping and early stopping. The finite rank-three solution with the lowest $\mathcal L(\mathbf Y)$ is centered and reported as $\hat{\mathbf X}$.

\section{Experimental Methodology}
\label{sec:experimental}

We evaluate 3D-PT through Monte Carlo numerical simulations implemented in MATLAB, using the IEEE 802.11n-inspired warehouse channel model described below. The reference warehouse is $140\times90\times14$~m with four 10-m-high rack structures, each adding 5~dB attenuation. For comparison, approximately 7.37~dB rack-induced loss at 2.4~GHz has been reported for 9-m racks \cite{Gong2018}. Nodes lie outside the racks at nominal heights $z\in\{1.2,3,5,8,11\}$~m; each nominal height is perturbed by independent zero-mean Gaussian jitter with standard deviation $0.35$~m to represent modest mounting-height variability rather than perfectly discrete height planes. Horizontal dimensions scale with $n$ to maintain approximately constant nominal density.

Packet detections follow a 2.412-GHz IEEE 802.11n-inspired model with 20-MHz bandwidth and MCS~0 \cite{Richter2019,Gonzalez2020}. The warehouse channel uses path-loss exponent 2.8, 6-dB shadowing, Rician fading, receiver variation, non-reciprocity, and 0.5-dB transmit-power error \cite{Rappaport1989UHF,Gong2016,Wu2012}. These quantities generate packet detections only; no localization method receives RSSI, true ranges, or environment geometry.

The small-network experiment uses $n=4,\ldots,15$ and 500 independent trials per configuration. The medium-network experiment uses $n\in\{10,20,30,40,50\}$ and 250 trials per configuration. Both compare $K\in\{1,4\}$; a sensitivity experiment fixes $n=10$, varies $K=1,\ldots,10$, and uses 500 trials per value. At $K=1$, 40 packets are sent at $+8$~dBm. At $K=4$, 40 packets are sent at each of $\{-10,-4,2,8\}$~dBm. For every PT configuration, MDS-MAP(P) and CATL-reproduction receive the same total packet budget at the common maximum power and reduce it to one reciprocal graph. Geometry and channels are paired across methods. Baseline comparisons are therefore equal-total-budget. PT $K=1$ versus PT $K>1$, however, compares complete per-level operating configurations rather than fixing the total number of observations.

\noindent\textbf{Baselines:}
We compare 3D-PT with MDS-MAP(P) \cite{Priyantha2005,Zhao2018}, a representative patch-based anchor-free connectivity method, and CATL \cite{Tan2013}, an anchor-free connectivity method designed for 3D irregular networks. MDS-MAP(P) uses two-hop local patches with a minimum four-node overlap for 3D patch merging; its parameters are fixed across all trials. CATL is a centralized reproduction of the mechanisms in \cite{Tan2013}, rather than the authors' original implementation, with fixed parameters taken from or motivated by that work. Neither baseline is tuned per trial. For each comparison, both baselines receive the same total packet budget as 3D-PT, concentrated at the common maximum transmit power, and reduce those observations to a single reciprocal connectivity graph.

For the $K$-sensitivity experiment, $K>1$ transmit-power levels are selected as approximately equally spaced settings from the fixed 1-dB grid spanning $-10$ to $+8$~dBm, with the lowest and highest levels fixed at $-10$ and $+8$~dBm; for $K=1$, only $+8$~dBm is used.

All simulations and runtime measurements were performed in MATLAB on an ASUS ROG Zephyrus G16 workstation equipped with an Intel Core Ultra 9 185H processor at 2.50~GHz and 32~GB RAM, running a 64-bit operating system.

\begin{table}[!t]
\centering
\caption{Fixed 3D-PT algorithm parameters.}
\label{tab:parameters}
\scriptsize
\renewcommand{\arraystretch}{1.06}
\begin{tabular*}{\columnwidth}{@{\extracolsep{\fill}} l c l c l c @{}}
\toprule
Parameter & Value & Parameter & Value & Parameter & Value\\
\midrule

$\theta_{\rm det}$ & 0.80 &
$\beta_r$ & 1.50 &
$\kappa_{\min}$ & 0.60\\

$\kappa_g$ & 0.55 &
$\mu_{\rm loc}$ & 0.08 &
$\chi_{\rm cc}$ & 0.25\\

$\chi_{\rm cf}$ & 0.30 &
$\gamma_{\rm loc}$ & 1.35 &
$\eta_{\rm loc}$ & 1.25\\

$\xi_{\rm strat}$ & 0.65 &
$\delta_g$ & 0.06 &
$\mu_g$ & 0.06\\

$\chi_g$ & 0.22 &
$\gamma_g$ & 1.25 &
$\tau$ & 0.15\\

$c_\rho$ & 2.50 &
$\lambda_G$ & 0.45 &
$\lambda_S$ & 0.002\\

\bottomrule
\end{tabular*}
\end{table}

For local candidates, the close set contains all $n-1$ ranked nodes when $n\leq10$; otherwise it contains $\min\{12,\max[8,\operatorname{round}(0.30(n-1))]\}$. The far set contains 0, 1, or $\min\{3,\operatorname{round}(0.10(n-1))\}$ nodes for $n\leq8$, $9\leq n\leq20$, or $n>20$, respectively. Stratification uses at most 12 uniformly spaced ranks after duplicate removal. Reliable unordered pairs generate global candidates at rank offsets $\{0.08,0.18,0.35,0.60\}$ of the reliable-pair count; at most eight comparisons per reliable pair are retained by weight.

All 3D-PT parameters in Table~\ref{tab:parameters} and the relation-count rules were selected during iterative development experiments using simulation realizations generated with random seeds distinct from those used for the held-out publication trials, and were frozen before the reported evaluation. The development and publication runs used the same warehouse and channel models but independent geometry and channel realizations.

The parameterization follows established principles from ordinal embedding, triplet learning, and robust ranking rather than previously prescribed numerical values. Prior work supports selecting and weighting ordinal constraints according to their informativeness or difficulty \cite{Haghiri2020,Tamuz2011,Jamieson2011}, while difficulty-dependent or adaptive-margin formulations provide precedent for increasing the requested separation for stronger or less ambiguous relations \cite{Zhang2019}. Soft ordinal stress functions with smooth, robust penalties are standard in ordinal embedding \cite{Terada2014}, and modern multidimensional scaling frameworks expose analogous weighting, robustness, and smoothing parameters as explicit tuning choices \cite{Mair2022}. Nonmetric formulations further establish that when only ordinal information is available, the recovered configuration is defined only up to scale, motivating explicit scale control and normalization \cite{Borg2005}.

The specific numerical values in Table~\ref{tab:parameters}, including the reliability thresholds, weighting exponents, margins, loss parameters, and objective weights, were selected during the development experiments and then fixed before evaluation. In particular, $\kappa_{\min}$ and $\kappa_g$ were chosen to suppress low-consistency relations while retaining useful local and global relation sets. Larger score gaps were given greater influence through the margin and weighting rules in \eqref{eq:localmargin}--\eqref{eq:localweight}, with the global relations treated analogously in \eqref{eq:globalmarginweight}. No parameters were changed after inspecting the reported publication-trial results.

\begin{figure*}[!t]
    \centering
    \subfloat[Small-network NRMSE, $K=4$.]{\includegraphics[width=0.315\textwidth]{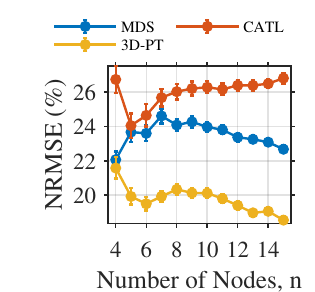}\label{fig:small_nrmse}}
    \hfill
    \subfloat[NRMSE versus $K$, $n=10$.]{\includegraphics[width=0.315\textwidth]{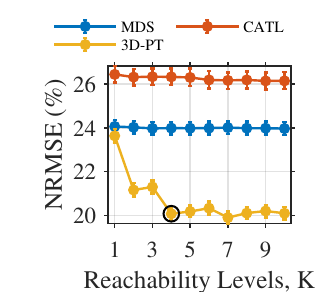}\label{fig:k_nrmse}}
    \hfill
    \subfloat[Medium-network NRMSE, $K=4$.]{\includegraphics[width=0.315\textwidth]{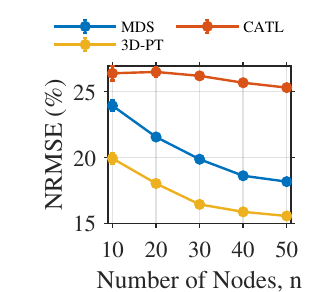}\label{fig:medium_nrmse}}
    \\[-1mm]
    \subfloat[Small-network runtime, $K=4$.]{\includegraphics[width=0.315\textwidth]{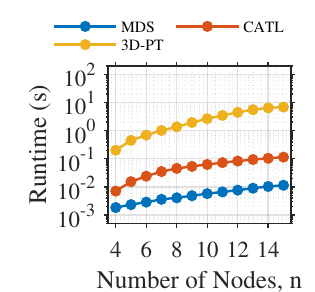}\label{fig:small_runtime}}
    \hfill
    \subfloat[Runtime versus $K$, $n=10$.]{\includegraphics[width=0.315\textwidth]{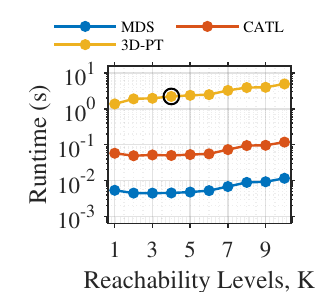}\label{fig:k_runtime}}
    \hfill
    \subfloat[Medium-network runtime, $K=4$.]{\includegraphics[width=0.315\textwidth]{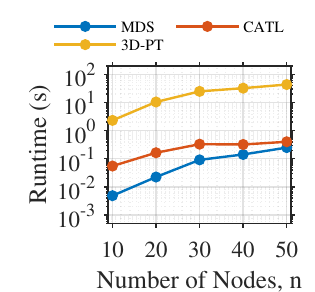}\label{fig:medium_runtime}}
    \caption{Warehouse accuracy and computation. NRMSE curves show means with 95\% percentile-bootstrap confidence intervals; some intervals, particularly in (c), are smaller than the plotting markers or line width and are therefore not visually distinguishable. Runtime is shown on a logarithmic scale. Baselines in (b) and (e) use equal-total packet observations for each $K$.}
    \label{fig:performance}
    \end{figure*}

Only trials with a connected reciprocal maximum-power graph are eligible for evaluation. For an eligible trial, a method is counted as successful if it reports successful reconstruction, returns a finite $n\times3$ coordinate matrix, and produces a finite aligned estimate. 
The success rate is the fraction of eligible connected trials satisfying these criteria. Each estimate is aligned to the true configuration using least-squares 3D Procrustes similarity alignment, allowing translation, rotation/reflection, and uniform scale \cite{Zhang2014, Flueratoru2020}. NRMSE is the 3D RMSE normalized by the true network diameter. For the NRMSE curves in Fig.~\ref{fig:performance}, mean NRMSE is reported over successful trials with 95\% percentile-bootstrap confidence intervals from 5000 resamples; the tables report the corresponding mean point estimates. Paired comparisons are restricted to trials in which both methods are successful and use a two-sided Wilcoxon signed-rank test with  $\alpha=0.05$. Runtime is the MATLAB wall-clock time of the localization routine itself; channel simulation, similarity alignment, metric calculation, and statistical post-processing are excluded.

    \section{Results and Discussion}
    \label{sec:results}
    
    \subsection{Small-Network Performance}
    
    \begin{table}[!t]
    \centering
    \caption{Warehouse small-network mean NRMSE (\%).}
    \label{tab:small}
    \scriptsize
    \setlength{\tabcolsep}{3pt}
    \renewcommand{\arraystretch}{1.06}
    \begin{tabular}{c c c c c}
    \toprule
    $n$ & PT $K=1$ & MDS-P & CATL-R & PT $K=4$\\
    \midrule
    4--6   & 25.43 & 23.12 & 25.15 & \textbf{20.34}\\
    7--10  & 24.56 & 24.24 & 26.05 & \textbf{20.13}\\
    11--15 & 23.10 & 23.24 & 26.45 & \textbf{19.16}\\
    \midrule
    4--15  & 24.17 & 23.54 & 25.99 & \textbf{19.78}\\
    \bottomrule
    \end{tabular}
    \end{table}

Across $n=4,\ldots,15$, PT $K=4$ achieves 19.78\% mean NRMSE, compared with 23.54\% for MDS-MAP(P), 25.99\% for CATL-reproduction, and 24.17\% for PT $K=1$ (Table~\ref{tab:small}). The corresponding descriptive reductions are 16.0\%, 23.9\%, and 18.2\%. For every $n=5,\ldots,15$, paired PT $K=4$ reductions are 10.44--17.74\% versus MDS-MAP(P) and 12.15--29.87\% versus CATL-reproduction; all are statistically significant.

The $n=4$ case requires separate interpretation. PT $K=4$ succeeds in 99.2\% of trials, whereas MDS-MAP(P) succeeds in 47.2\%. PT has the lower overall successful-trial mean NRMSE (21.58\% versus 22.07\%), but the comparison restricted to the 227 common-success trials favors MDS-MAP(P), illustrating a strong common-success selection effect at the minimum network size. PT $K=4$ succeeds in every connected trial for $n=5$--15; MDS-MAP(P) reaches 100\% success from $n=6$ onward, while CATL-reproduction remains less reliable in the smallest networks.

\subsection{Effect of the Number of Power Levels}

At $n=10$, mean PT NRMSE decreases from 23.64\% at $K=1$ to 20.07\% at $K=4$, a 15.1\% reduction. It then remains within 19.90--20.33\% for $K=4$--10; the minimum, 19.90\% at $K=7$, has an overlapping confidence interval with $K=4$. Mean runtime increases from 1.37~s at $K=1$ to 2.21~s at $K=4$ and 4.95~s at $K=10$, while packet count grows from 40 to 160 and 400. Four levels therefore capture most of the available ordinal benefit at substantially lower observation and computational cost than denser power sweeps. Because 40 packets are collected per active level, this experiment characterizes complete operating configurations rather than isolating $K$ under a fixed total packet count.

\subsection{Medium-Network Performance}

\begin{table}[!t]
\centering
\caption{Warehouse medium-network results at $K=4$.}
\label{tab:medium}
\scriptsize
\setlength{\tabcolsep}{2.6pt}
\renewcommand{\arraystretch}{1.04}
\begin{tabular}{c c c c c}
\toprule
$n$ & MDS-P & CATL-R & 3D-PT & $K=1\!\rightarrow\!4$\\
 & \multicolumn{3}{c}{NRMSE (\%)} & reduction (\%)\\
\midrule
10 & 23.94 & 26.40 & \textbf{19.94} & 15.26\\
20 & 21.57 & 26.50 & \textbf{18.04} & 19.07\\
30 & 19.88 & 26.21 & \textbf{16.45} & 21.18\\
40 & 18.63 & 25.68 & \textbf{15.89} & 18.43\\
50 & 18.18 & 25.32 & \textbf{15.57} & 16.83\\
\bottomrule
\end{tabular}
\end{table}

PT $K=4$ has the lowest mean NRMSE at every tested medium-network size, decreasing from 19.94\% at $n=10$ to 15.57\% at $n=50$ (Table~\ref{tab:medium}). Averaged across sizes, its 17.18\% NRMSE is 16.0\% below MDS-MAP(P), 34.0\% below CATL-reproduction, and 18.1\% below PT $K=1$. Both PT $K=4$ and MDS-MAP(P) succeed in all 1250 connected trials; CATL-reproduction succeeds in 97.84\%. The principal medium-network advantage is therefore improved accuracy rather than success rate.

The accuracy gain requires more computation. Mean PT runtime grows from 2.31~s at $n=10$ to 43.3~s at $n=50$, compared with 0.0049--0.249~s for MDS-MAP(P) and 0.055--0.401~s for CATL-reproduction. The present MATLAB implementation prioritizes reconstruction quality; reducing candidate-relation count and multi-restart optimization cost is an important implementation direction.

\subsection{Interpretation and Limitations}

The results indicate that the gain is not obtained merely by repeating maximum-power connectivity measurements. MDS-MAP(P) and CATL-reproduction receive the same total packet count as PT for each baseline comparison, but their observations are collapsed to one reciprocal graph. In contrast, PT preserves where each pair changes state along the ordered power ladder. This separates pairs that are indistinguishable in the maximum-power graph and supplies additional local orderings; the global pair comparisons then discourage configurations that satisfy many reference-centered relations while remaining distorted at larger scale. The benefit is largest relative to CATL-reproduction, whose hop-based construction is more sensitive to sparse or nonuniform connectivity.

The $K$ sweep also identifies an operational limit. Once most useful pairs have distinct monotone signatures, adding levels mainly subdivides already established transitions. The resulting score gaps and selected relation set change little, explaining the narrow NRMSE range for $K=4$--10 despite increasing packet count and runtime. This observation supports $K=4$ for the evaluated power range, but it is not a universal optimum: different radios, power spacing, receiver thresholds, or propagation conditions may shift the saturation point.

Several limitations bound the conclusions. First, PT $K=1$ versus PT $K=4$ is a comparison of complete per-level configurations, not a fixed-total-packet ablation; it therefore measures the practical benefit of the multi-level configuration rather than attributing the entire gain solely to the number of levels. Second, all results are simulation based and depend on the stated warehouse and channel model, although parameter tuning and publication trials are separated. Third, 3D-PT returns relative geometry and requires a connected maximum-power graph; absolute coordinates or disconnected-component alignment would require additional information. Finally, the non-convex multi-restart solver is substantially slower than the baselines.

\section{Conclusion}
\label{sec:conclusion}

3D-PT reconstructs relative 3D geometry from multi-level binary reachability without RSSI magnitudes, metric ranges, surveyed anchors, or prior coordinates. In the evaluated warehouse, four power levels improve mean NRMSE over single-level PT and both connectivity baselines, with gains persisting from very small to 50-node networks. Additional levels provide little accuracy beyond $K=4$ but continue to increase packet and computational cost. These results support multi-level reachability as a useful source of ordinal geometry, while physical validation and faster optimization remain necessary next steps.

\section*{Acknowledgment}
This research was partially funded by the McGill University Vadasz Scholars Program and by the Natural Sciences and Engineering Research Council of Canada (NSERC), [reference number 260250]. Cette recherche a été partiellement financée par le Conseil de recherches en sciences naturelles et en génie du Canada (CRSNG), [numéro de référence 260250]. Ce projet de recherche n\textsuperscript{o}~324302 est rendu possible grâce au financement du Fonds de recherche du Québec.

\bibliographystyle{IEEEtran}
\bibliography{references}

\end{document}